\documentclass{aa}  

\usepackage{graphicx,url}
\usepackage[dvipsnames]{xcolor}
\usepackage[normalem]{ulem}
\usepackage{mathrsfs,amssymb,amsmath}
\usepackage[varg]{txfonts}
\usepackage{booktabs}
\usepackage{multirow}
\usepackage{lineno}
\usepackage{siunitx} 
\usepackage[breaklinks, colorlinks, citecolor=blue]{hyperref}

\newcommand{\mJypbm}{\text{mJy}\,\text{beam}$^{-1}$}

\begin{document} 
   \title{A broadband view of the thermal and non-thermal emission from the embedded massive star cluster RCW~38}


   \author{G. Peron
          \inst{1}
          \and
          A. Bracco \inst{2,1,3}
          \and
          G. Morlino \inst{1}
          \and
          S. Mantovanini \inst{4}
          \and
          E. Amato \inst{1}
          \and
          D. Galli \inst{1}
          \and
          M. Padovani \inst{1}
          }

   \institute{INAF Osservatorio Astrofisico di Arcetri, Largo Enrico Fermi, 5, 50125, Florence, Italy\\
             \email{giada.peron@inaf.it}
             \and
             LUX, Observatoire de Paris, Université PSL, Sorbonne Université, CNRS, 75014 Paris, France
             \and
Laboratoire de Physique de l'Ecole Normale Sup\'erieure, ENS, Universit\'e PSL, CNRS, Sorbonne Universit\'e, Universit\'e de Paris, F-75005 Paris, France
\and
Max-Planck-Institut f\"{u}r extraterrestrische Physik, Gie{\ss}enbachstra{\ss}e 1, D-85748 Garching, Germany
}
  \abstract
   {Gamma-ray emission has now been detected from a variety of source classes in the Galaxy, including clusters of young massive stars. RCW~38, a very young embedded massive star cluster, is a case of particular interest: its gamma-ray emission detected up to hundreds of GeV, provided the first observational evidence of high-energy particle acceleration powered exclusively by stellar winds.}
   {In this work, we aim to characterize the emission mechanisms responsible for the gamma-ray flux in RCW~38 and to provide estimates of the acceleration efficiency, as well as the fraction of accelerated electrons compared to protons, $K_{\rm ep}$.}
  {We present the most comprehensive multi-wavelength study of a single star cluster to date. Our analysis ranges from MHz radio observations obtained with the GLEAM-X survey from the Murchison Widefield Array (MWA) to GeV gamma-ray data from {\it Fermi}-LAT, and includes GHz and THz measurements from Parkes, {\it Planck}, and IRAS. We model the thermal and non-thermal emission of RCW~38 —- including synchrotron, free-free, dust, bremsstrahlung, pion-decay, and inverse Compton processes -— using an eight-parameter model constrained by the Markov chain Monte Carlo method.}
   {Our results support an interpretation in which the gamma-ray emission from RCW~38 is produced by hadronic interactions with the host molecular cloud. We derive robust constraints on the electron-to-proton ratio, with $K_{\rm ep}\lesssim 10^{-3}$, and on the acceleration efficiency, estimated to be $\gtrsim 1\%$, consistent with the values required to explain the cosmic-ray composition, and in particular its $^{22}$Ne anomaly.}
   {These results strengthen the idea that stellar clusters play a significant role as contributors to cosmic-ray protons in our Galaxy at least up to energies of a few TeV. Future investigations with the next generation of ground-based detectors will determine whether they also play a relevant role at higher energies, particularly in the context of the cosmic-ray knee.}
   {}


   \maketitle

\section{Introduction}
Clusters of young massive stars have recently been confirmed as sources of high-energy particles and consequently of non-thermal emission, even in the earliest phase of their life \citep[< 2 Myr;][]{Peron2024ThePopulation, Peron2025, Peron2025icrc}, before hosting supernova events and their remnants (SNRs). At such a young age, stellar clusters (hereafter, SCs) are still embedded in the gas cocoon in which they formed, and reveal themselves through the feedback emission into the surrounding gas. H\textsc{ii} regions are commonly used to identify young stellar systems in the radio, optical and near-infrared emission. It is interesting to note that a strong correlation has recently emerged between H\textsc{ii} regions and unidentified {\it Fermi}-LAT sources \citep{Peron2024On}, supporting the idea that SCs are important sources of accelerated particles, at least up to TeV energies. Particles in these environments could be energized in different ways: via diffusive shock acceleration (DSA) at the wind-wind collision shock of binary stellar systems \citep{Reimer2006} or at the termination shock of the collective wind \citep{Morlino2021}; or via the second-order Fermi mechanism, induced by the scattering of the particles on the turbulent magnetic fields found in the wind blown bubble \citep{Vink2024}. In all cases,  particles are accelerated in the inner regions and then propagate through a low density cavity---the wind-inflated bubble---until they reach the dense swept-up material of the shell, where the particles can interact effectively and produce gamma-ray emission through inelastic nuclear collisions or bremsstrahlung.  
In regions of high gas density, the dominant mechanism of non-thermal emission is determined by the fraction, $K_{\rm ep}$ (see definition in Eq.~\ref{eq:Ne}), of accelerated electrons relative to protons. This quantity has been constrained observationally in a few SNRs, including  both young SNRs like RX~J1713.7-3946 \citep{BerezhkoVolkRXJ1713}, Vela Jr. \citep{Berezhko2009}, Tycho \citep{MorlinoCaprioli}, SN1006 \citep{Morlino2010}, founding a $K_{\rm ep}$ between 10$^{-4}$ and 10$^{-3}$ , and middle aged SNRs like W44, W28 \citep{Zirakashvili2017},  and Cygnus Loop \citep{Loru2020}, showing  $K_{\rm ep}$ ranging from $\sim5\times 10^{-3}$ to 0.15. Meanwhile, to explain the observed ratio of CR electrons over CR protons measured at Earth, accounting for the different energy losses of these two populations during their propagation in a state-of-the-art diffusion scenario \citep[e.g.][]{Evoli2020}, an injected $K_{\rm ep}$ of $\sim 10^{-3}-10^{-2}$  is needed. Interestingly, recent magneto-hydrodynamic (MHD) simulations of the propagation of CR electrons and protons through the multi-phase ISM, are able to reproduce the near-Earth CR spectrum if at injection a value of $K_{\rm ep}\sim0.05$ is assumed \citep{Linzer2025}.  It stands out clear that this quantity is key to understand the sources of Galactic CRs. Still, $K_{\rm ep}$ remains unknown for stellar systems, preventing us for a clear assessment of their contribution to the bulk of CRs in the Galaxy.
While protons mainly produce emission in the gamma-ray energy band, electrons are expected to emit also at lower energies. In addition to non-thermal bremsstrahlung, which is supposed to peak at around 1~MeV\footnote{{In this treatment we neglect electron energy losses due to ionization, which may steepen the spectrum in the energy range from 1 to 100~MeV \citep{PadovaniLosses2024}. This should not cause any concern as we do not have any data in that energy range.}}, synchrotron emission is expected to span a broad energy range from radio to X-rays. Hints of synchrotron emission, namely of soft power-law radio spectra, have been detected in a handful of Galactic  H\textsc{ii} regions \citep{Mucke2002,Veena2016,Nandakumar2016,Meng2019} including the gamma-ray-emitting region NGC~3603.  The radio emission detected in these objects is interpreted in terms of local electron sources \citep{Padovani2019Non-thermalRegions}.

In this paper, we examine the case of RCW~38, an H\textsc{ii} region embedded in a very young \citep[0.5 Myr;][]{Wolk2006} star cluster with $\sim 3000$ stars, 30 of which are early-type O stars. RCW~38 has been detected at high energies with {\it Fermi}-LAT, revealing a population of protons up to a few TeV \citep{Peron2024ThePopulation}. This source is also prominent at other wavelengths, from radio \citep{Tremblay2022} to X-rays \citep{Wolk2006}, with the emission identified as free-free and thermal radiation in these wavebands.  We investigate further the origin of this emission in light of recent radio, gamma- and X-ray observations, evaluating the contribution of synchrotron emission expected in the radio band, once constraints from higher energies are taken into account.

The paper is structured as follows: in Sect.~\ref{sec:obs}, we examine the available observational data relating to the source and briefly describe the data analysis methods; in Sect.~\ref{sec:model}, we present our model for the population of primary protons and electrons. In the same Section, we present the parameters used for the radiative models. We discuss the results and their implications in Sect.~\ref{sec:result}, while in Sect.~\ref{sec:conclusion} we draw our conclusions. The paper contains three appendices that complement the text by adding some details on the modeling and on the data analysis.

\section{Available observations and methods}
\label{sec:obs}

In this section we describe the data and the methods used for the analysis. 

\subsection{Gamma-ray data}
An extended gamma-ray emission spanning a circular area of radius $\theta_{\gamma}= 0.21^\circ \pm 0.04^\circ$ -- corresponding to a physical radius $R_{\gamma}\sim 6~\mathrm{pc}~(d/1600~\mathrm{pc})$ at the given distance $d$-- has been recorded at the location of RCW~38 by the analysis of data collected by the LAT instrument on-board the {\it Fermi} satellite \citep{Peron2024ThePopulation}. The morphology of the gamma-ray emission is represented in Fig. \ref{fig:map} by white contours over-plotted to the map of radio data at three different frequency (as described later);  the shape of the gamma-ray emission well matches the emission of warm dust, represented as grey contours in the same figure. 

The spectral energy distribution (SED) has been recorded in the energy range from 500~MeV to $\lesssim$ 1 TeV, considering 6 equally spaced logarithmic bins, and is modeled as a power-law function of energy:
\begin{equation}
    F(E_{\gamma})=F_0 \bigg(\frac{E}{E_0}\bigg)^{-\Gamma},
\end{equation}
with a power-law index, $\Gamma=2.56\pm0.05$ and a flux normalization at $E_0=1~\mathrm{GeV}$ of $F_0=3.9\pm0.2 ~\mathrm{MeV^{-1}cm^{-2}s^{-1}}$. The total luminosity derived from {\it Fermi}-LAT observations is $L_\gamma=5\times 10^{33}$ erg~s$^{-1}$. {Later studies investigated the same region finding compatible results in terms of gamma-ray flux and morphology \citep{Pandey2024,Ge2024RCW38}. Systematic differences among these works are briefly discussed in Appendix \ref{sec: FermiCFR}}.

\subsection{X-ray data}
RCW~38 was studied in the X-ray domain (between 1 and 10 keV) first by Chandra \citep{Wolk2002Discovery38} and subsequently by Suzaku \citep{Fukushima2023}. Both studies revealed an extended diffuse emission that was divided into an inner ($<2^\prime$) and an outer region ($2^\prime-5.5^\prime$). Two explanations have been proposed for the emission in both regions: one requiring a non-thermal component in addition to a thermal component, another requiring two thermal components of different temperatures (of $\sim 1$~keV and $\sim 6$~keV). It has been speculated that the X-ray non-thermal component seen by Chandra and Susaku may arise from synchrotron emission produced by multi-TeV electrons, although no explanation is provided for the origin of such high-energy electrons. The possible synchrotron origin of this component is discussed in Appendix~\ref{sec: syncX}, where we assess whether electrons accelerated by the winds of RCW~38 could be responsible for this emission. In the following, we assume the X-ray emission to be thermal, and, therefore, the X-ray flux is used as an upper limit on any additional contribution from our model in that energy range.

\subsection{Far-infrared and sub-mm data}
In the far-infrared (FIR) and sub-mm domains, we used publicly available total-intensity data from the IRAS\footnote{\url{https://lambda.gsfc.nasa.gov/product/foreground/fg_iris_info.html}} \citep{mamd2005} and {\it Planck}\footnote{\url{https://pla.esac.esa.int/}} \citep{planckI2020} satellites. We included IRAS bands 3 and 4 at 60~$\mu$m and 100~$\mu$m, respectively, and all {\it Planck} frequency channels between 30\,GHz and 857\,GHz.

Since RCW~38 is a bright source located in the Galactic plane, we did not apply component-separation algorithms to remove extragalactic contributions such as the cosmic microwave background (CMB) and the cosmic infrared background (CIB). At the scale of RCW 38 and in this region of the sky, the emission is expected to be dominated by Galactic foregrounds, i.e. dust, free-free, and synchrotron. 

RCW~38 has also been identified as not emitting anomalous microwave emission \citep[AME;][]{planckXV2014}. We therefore neglected AME, CIB, and CMB contributions in the flux modelling of the source in the FIR/sub-mm bands.

\subsection{Radio data}
\begin{figure}
    \centering
    \includegraphics[width=1\linewidth]{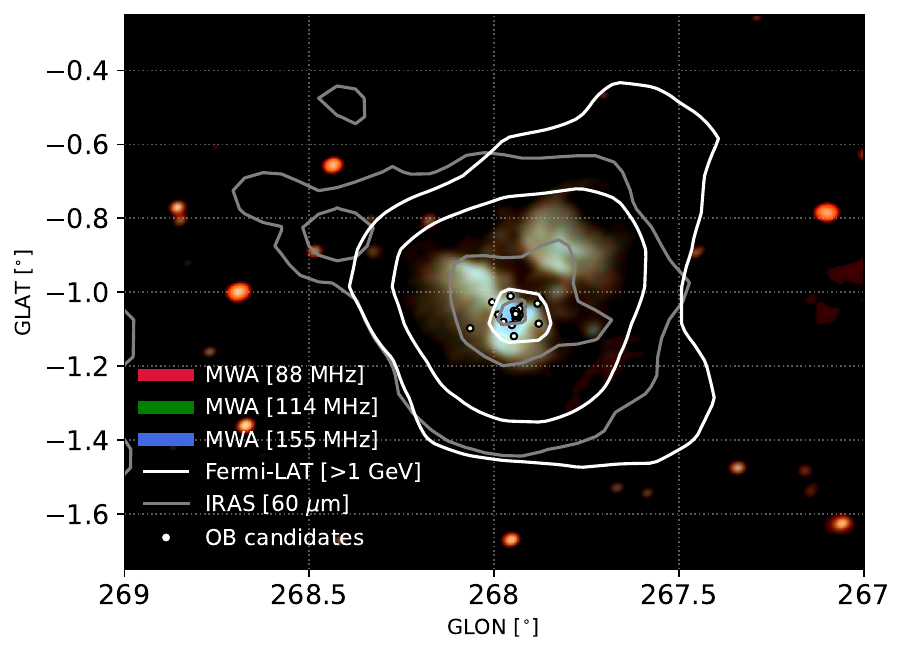}
    \caption{RGB map of RCW~38 as detected by MWA within the GLEAM-X survey \citep{Mantovanini2025}, in three different bands: 88 MHz (red), 114 MHz (green) and 155 MHz (blue). The grey contours represent emission unveiled by IRAS at 60 $\mu$m, while the white contours is the gamma-ray signal derived by {\it Fermi}-LAT above 1 GeV. In the latter case, the contours indicate significance level corresponding to a test-statistics of 50,100, and 500. }
    \label{fig:map}
\end{figure}

In the radio range, we considered both centimeter- and meter-wavelength data. In the former case, we used the publicly available\footnote{\url{https://sites.google.com/inaf.it/spass}} total-intensity map at 2.3\,GHz from the S-band Polarization All-Sky Survey \citep[S-PASS;][]{Carretti2019}. The S-PASS total-intensity map, obtained with the 64-m dish of the Parkes-Murriyang telescope, was zero-level calibrated using absolute flux measurements available in the literature.
At meter wavelengths, we used the recently released Galactic Plane (GP) data \citep{Mantovanini2025} obtained with the Murchison Widefield Array \citep[MWA,][]{Tingay2013} as part of the Galactic and Extragalactic All-Sky MWA eXtended survey \citep[GLEAM-X,][]{Hurley2022,Ross2024}, which covers the southern GP ($\ang{233} < l < \ang{44}$ and $|b| < \ang{11}$) over a frequency range of $72 - 231$\,MHz divided into 5 wide sub-bands ($\approx 30$\,MHz) and 20 further narrow sub-bands ($\approx 8$\,MHz). The resulting Stokes $I$ mosaics were obtained by combining observations taken with the MWA in its extended Phase-II configuration \citep{Wayth2018}, featuring baselines from $12$~m to $5$~km, and observations from the original GLEAM survey \citep{Wayth2015} to enhance sensitivity over a broad range of spatial scales ($45^{\prime\prime} - \ang{15}$), and achieve noise levels of $\approx 3-6$~\mJypbm. For our analysis, we used the total-intensity mosaics in the three lowest wide sub-bands, centered at $88$\,MHz, $118$\,MHz, and $155$\,MHz, respectively. In Fig.~\ref{fig:map}, an RGB image of RCW~38, constructed from the three MWA frequencies and overlaid with gamma-ray and IRAS band 3 contours, is shown together with the positions of the OB candidates. 

\subsection{Data processing: smoothing and background subtraction}\label{ssec:methods}
To  compute the flux of RCW~38 consistently across the entire range of frequency bands, we smoothed  all {the low-energy} maps to a common full width at half maximum (FWHM) resolution. In the case of IRAS, {\it Planck}, and S-PASS, we used maps in {\tt HEALPix}\footnote{\url{http://HEALPix.sf.net}} format~\citep{Gorski2005} and processed them with the {\tt healpy} Python package~\citep{Zonca2019}. For each band $i$, the convolution kernel was chosen such that the final FWHM satisfied
\begin{equation}
\theta_{\mathrm{kernel},i} = \sqrt{\theta^2 - \theta_{0,i}^2},
\end{equation}
where $\theta_{0,i}$ represents the original FWHM of the $i$-th band and $\theta=\theta_{\gamma}$ is the adopted common resolution. The main characteristics of the datasets are summarized in Table~\ref{tab:maps}.

No smoothing was applied to the 30\,GHz and 44\,GHz bands, since in these cases $\theta_{0,i} > \theta$. Instead, we applied empirical aperture-correction factors of 1.8 and 1.5 at 30\,GHz and 44\,GHz, respectively. These factors were derived by estimating the flux loss within a circular aperture of radius $\theta$ after convolving several reference templates --described in detail in the following -- at the corresponding instrumental resolutions $\theta_{0,i}$. The correction factors were computed using independent templates, including S-PASS 2.3\,GHz, {\it Planck} 353\,GHz, and IRAS 100\,$\mu$m maps. All templates yielded consistent correction values.

Following convolution at the common angular resolution, we performed a local background subtraction. As described in Appendix~\ref{sec:appendix_radio}, the background level was estimated from the mean flux measured in an external annular region between $0.3^{\circ}$ and $3^{\circ}$ from RCW~38.

The integrated flux in each frequency band was then computed (in units of erg\,cm$^{-2}$\,s$^{-1}$) by summing the background-subtracted specific intensity over all pixels contained within a circular aperture of radius $\theta$ centered on RCW~38, corresponding to the extension of the gamma-ray signal \citep{Peron2024ThePopulation}. Since the aperture size is significantly larger than the pixel scale in all bands, the original pixel grids were retained without introducing flux losses. The relative error on the integrated flux is dominated by systematic effects and background subtraction at a level of 15~\%. 
  
\section{Emission modeling}
\label{sec:model}

In this Section, we describe the initial assumptions about the ambient parameters and particle distribution adopted for modeling the different emission components that we use to fit the available data. Only primary electrons and protons were considered, while we assessed the level of secondary electrons only after the first fit, finding it to be negligible. The best-fit parameters were derived using a Markov Chain Monte Carlo \citep[MCMC;][]{emcee2013} method, using 128 walkers running for 3000 iterations to efficiently sample the parameter space within reasonable computation time. A summary of the parameters, together with their boundaries can be found in Table \ref{tab:params}; the rationale for choosing these values is given in the following sections.

\subsection{Primary particles} 
\label{subsec:primaries}
We assume that the gamma-ray emission is produced by two populations of relativistic particles, namely electrons and protons. The spectral distribution of protons is described, as a function of the particle momentum, $p$, by a power-law with an exponential cutoff:  
\begin{equation}
\label{eq:Np}
N_p(p) = N_{0,p} \bigg(\frac{p}{p_0} \bigg)^{-\alpha_p} \exp(-p/p_{p,{\rm max}}),
\end{equation}
where $N_{0,p}$ is the normalization at the pivot value $p_0c\approx 10$~GeV and is derived in our fit to the data. The spectral index $\alpha_p$, is fixed by gamma-ray observations to the value of 4.6 \citep{Peron2024ThePopulation}. The latter could be produced either by an intrinsically steep injection spectrum, or by the combination of a hard injection spectrum and energy-dependent escape, as in the case of fast diffusion. The first scenario may occur in the case of a relatively weak shock, with a small Mach number ${\cal M}$. An estimate of the Mach number can be obtained using the wind-bubble model by \cite{Chevalier-Clegg:1985} if the cluster radius, $R_{\rm cl}$, and the shock radius, $R_{\rm ts}$, are known. The relation between the three reads:

\begin{equation}
    \mathcal{M}^{\frac{2}{(\gamma-1)}} \, \left( \frac{\gamma-1+2/\mathcal{M}}{1+\gamma} \right)^{\frac{\gamma+1}{2 (\gamma-1)}} = \left( \frac{R_{\rm ts}
    }{R_{\rm cl}}\right)^2 \,,
\end{equation}
where $\gamma=5/3$ is the fluid adiabatic index. The size of the wind termination shock, $R_{\rm ts}$, is computed by balancing the ram pressure of the wind against the gas pressure in the surrounding bubble, and eventually depends on the cluster wind speed and density. As a reference, we consider the recent estimates provided by \cite{Pandey2024} for the collective wind luminosity and wind speed, $L_{\rm w}=8\times 10^{36}$~erg~s$^{-1}$ and $V_{\rm w}$=2600~km~s$^{-1}$, which yield to a mass-loss-rate of $\dot{M}\sim 3.8\times 10^{-6}\ \mathrm{M_{\odot}~yr^{-1}}$. Therefore we can estimate $R_{\rm ts}$ as:

\begin{equation}
\begin{split}
R_{\rm ts} = 1.5~\eta^{-1/5}_M
&\left(\frac{\dot{M}}{3.8\times 10^{-6}\,\mathrm{M}_{\odot}\,\mathrm{yr}^{-1}}\right)^{3/10}
\left(\frac{V_{\rm w}}{2600\,\mathrm{km}\,\mathrm{s}^{-1}}\right)^{1/10} \\
&\times
\left(\frac{n}{100\,\mathrm{cm}^{-3}}\right)^{-3/10}
\left(\frac{T_{\rm age}}{0.5\,\mathrm{Myr}}\right)^{2/5}
\,\mathrm{pc} .
\end{split}
\end{equation}
$R_{\rm ts}$ ranges from 1.5 to 0.5 pc, with the external density varying from 100 to 1000 cm$^{-3}$; these values of the density are in agreement with high resolution observations of CO line emission in RCW 38 \citep{Fukui2017Correspondence}. The factor $\eta_M$ represents the fraction of wind energy that is converted into mechanical energy and is generally much smaller than unity; in the following we assume $\eta_M=1$, but note that {this implies that our estimate of $R_{\rm ts}$ should be regarded as an upper limit and may change by a factor of a few} due to this effect. The effective cluster size is not known, but assuming a reasonable size of $\simeq 1$ pc \citep{Tarricq2022,Just2023}, naturally gives $\mathcal{M} \simeq 2.6$ and hence an injection spectrum of $\alpha = 3\sigma/(\sigma-1)\simeq 4.6$    
in good agreement with the value inferred from observations\footnote{$\sigma$ is the shock compression ratio, linked to the Mach number through: $\sigma= (\gamma+1) \mathcal{M}^2/((\gamma-1)\mathcal{M}^2+2)$, where $\gamma=5/3$ is the adiabatic index.}. A scenario in which the observed spectrum is the same as the injection spectrum would have interesting consequences on particle transport within the bubble: in the entire range of particle energies ($\sim$ 10 GeV-1 TeV) probed by our gamma-ray data, the condition $T_{\rm adv} < T_{\rm diff}$ must then hold, implying a diffusion coefficient $D < R^2_{\gamma}/6 T_{\rm adv} \sim 10^{25} {\rm cm^{2} s^{-1}}$, when assuming that $T_{\rm adv}$ is a fraction $\sim 0.5$ of the age of the system \citep{Weaver1977InterstellarEvolution.}. To understand whether such assumption is realistic, we need to estimate the magnetic field that characterizes the region. The latter can be calculated equating the magnetic energy flux
($\propto R_{\rm ts}^2V_{\rm w}B^{2}$) to a fraction $\eta_B$ of $L_{\rm w}$. The value of $\eta_B$ cannot be constrained with available information, although it is commonly assumed to be a small fraction \citep[$\eta_B\leq 10$\%][]{Badmaev2022InsideSimulations}  of the wind power. 
Using  $\eta_B=0.1$, we obtain $B_{\rm ts}\sim 10~\mu$G for the given values of $R_{\rm ts}$ and $L_{\rm w}$, and hence a compressed magnetic field downstream of the shock $B_2\sim \sqrt{11} B_{\rm ts}\sim 40\mu$G. In such a magnetic field, a slow diffusion coefficient is compatible with a Bohm-like scenario, but can be obtained also considering Kolmogorov or Kraichnan turbulence, if the coherence length of the magnetic field is short ($L_c< 0.01$ pc), due for example to the corrugation of the shock and/or to the presence of clumps. Being the system embedded in a dense \citep{Fukui2016} and clumpy medium, where dense clumps of sub-parsec scale are observed \citep{Torii2021alma}, we consider this hypothesis reasonable and assume, in the following discussion, that the spectrum of particles is intrinsically steep. 
We discuss the consequences of a scenario in which diffusion steepens the spectrum later in the article.

\begin{figure}
    \centering
    \includegraphics[width=0.95\linewidth]{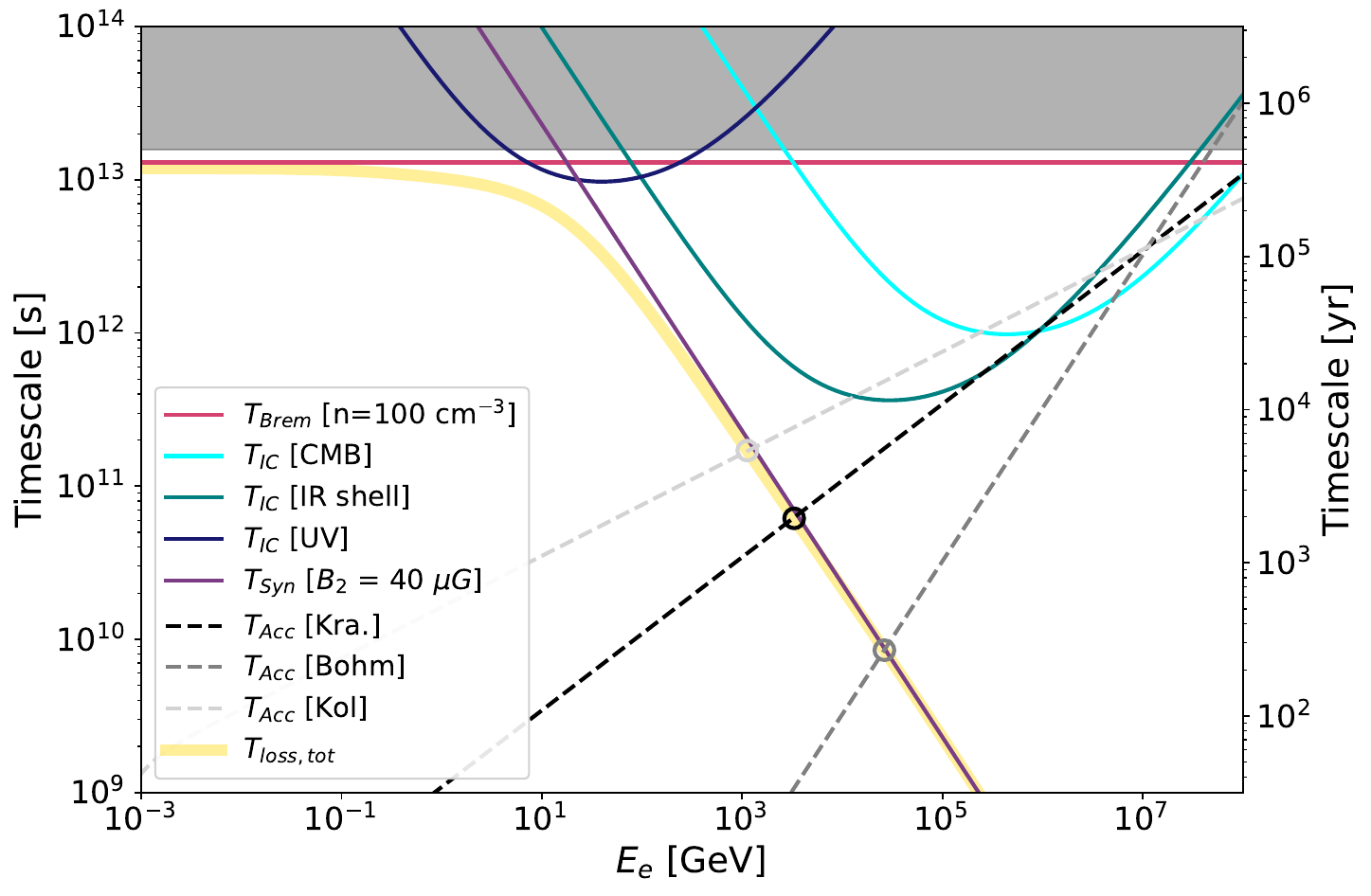}
    \caption{Energy-loss and acceleration timescales of electrons are compared. The latter are calculated considering the diffusion coefficient in three different turbulence scenarios, Bohm, Kolmogorov, and Kraichnan, as indicated in the figure legend. The open circles indicate, for each case, the maximum energy, defined by the points at which the timescales intercept. The grey area indicates an epoch longer than the age of RCW~38, and therefore represents a forbidden area in this plot.   }
    \label{fig:eloss}
\end{figure}

Differently from the spectral slope, the maximum momentum, $p_{p,{\rm max}}$, does not appear clearly from the data: the spectrum extracted from {\it Fermi}-LAT observations is well modeled with a power-law with no clear bending. An estimate, however, can be derived imposing: 

\begin{equation}
  D(E_{\max})/V_{\rm w} = R_{\rm ts},
\end{equation}
where $R_{\rm ts}$ is considered to be the size of the acceleration region \citep[as e.g. in][]{Morlino2021}. We stress that the other mechanisms, like turbulent acceleration and acceleration at colliding winds of binary stars, usually predict a smaller or at most comparable value for the maximum energy, but are here neglected because they are expected to be subdominant in terms of total available power. 
We refer to the recent work by \cite{Menchiari2024} (Eq. 22-24) for useful expressions for the maximum energy, considering the different energy dependence of the diffusion coefficient. Using the upper limit on the diffusion coefficient discussed above, the maximum energy, $E_{p,\max}\sim p_{p,\max}c$ ranges from 10~TeV to 1~PeV. All values in this range are compatible with the available data, therefore we fix the maximum energy of protons to 100~TeV. We discuss possible strategies for constraining the maximum energy of protons exploiting the upcoming ground-based facilities for very-high-energy gamma rays in Sect.~\ref{sec:conclusion} below.
Unlike protons, the spectrum of electrons is heavily influenced by energy losses, which determine both the maximum energy and the slope. We write the spectral distribution of electrons as: 
\begin{equation}
\label{eq:Ne}
N_e(p) = K_{\rm ep} N_{0,p} \frac{\min[T_{\rm loss},T_{\rm age}]}{T_{\rm age}}\bigg(\frac{p}{p_0} \bigg)^{-\alpha_p}\exp(-p/p_{e,{\rm max}}).
\end{equation}
$K_{\rm ep}$ represents the ratio between electrons and protons and is one of the parameters derived from our fit. $T_{\rm loss}$ is the timescale of energy losses: at the termination shock, the dominant energy-loss mechanisms are synchrotron and inverse Compton (IC) scattering on the ambient radiation fields. The timescales for synchrotron losses, calculated for the magnetic field, $B_2$, downstream of the shock and for IC scattering on photons from the CMB, the infrared thermal emission, and the stellar photon fields are shown in Fig. \ref{fig:eloss}. The latter two contributions amount to $\sim 10$~eV~cm$^{-3}$ and to $\sim 280$~eV~cm$^{-3}$, respectively, and have been derived from observations of the thermal dust emission made with IRAS (and discussed in Sect.~\ref{sec:thermal}) and of the stellar population made with Chandra \citep{Wolk2006}. The maximum electron energy is eventually obtained by comparing the energy-loss and acceleration timescales: $T_{\rm acc}\sim D(E)/(8V^2_{\rm w})$. Different diffusion coefficients yield different $T_{\rm acc}$: in Fig. \ref{fig:eloss} we show the case for Bohm, Kraichnan, and Kolmogorov diffusion in a magnetic field of strength $B_{\rm ts}$.  As expected, the choice of the diffusion coefficient strongly affects $E_{e,{\rm max}}$, which ultimately varies from a few TeV to 100 TeV. In this case too, the corresponding energy for gamma-rays falls beyond the interval probed by {\it Fermi}-LAT and therefore a proper constraint can only be derived in the future. Similarly to what done for protons, we therefore fix the maximum energy of electrons to 10 TeV.

\subsection{Thermal emission}
\label{sec:thermal}
In the following, we describe the thermal components that we use to model the emission. As is typical for H\textsc{ii} regions, we considered a component of free-free emission, and the thermal emission of dust, peaked in the FIR.

\subsubsection{Dust thermal radiation}
The thermal emission of dust can be described by a modified black-body function, as
\begin{equation}\label{eq:BB}
    I^{\rm d}_\nu \propto N_{{\rm H}_2} B_{\nu}(\nu,T_{\rm d}) \, \nu^{\beta},
\end{equation}
where $B_{\nu}$ is the emission spectrum of a black-body of temperature $T_{\rm d}$. Typically, $\beta$ ranges from 1 to 2.
This emission component depends on the gas column density $N_{{\rm H}_2}$ and the dust temperature $T_{\rm d}$. The analysis of \cite{Smith1999} of the mid-infrared emission revealed regions with different dust temperatures and column densities, ranging from $\sim 30$~K to $\sim 170$~K, and from $10^{20}$~cm$^{-2}$ to $10^{22}$~cm$^{-2}$, respectively. Using different tracers, \cite{Bourke2001NewClouds} and \cite{Wolk2002Discovery38} also found column density values in this range, while \cite{Fukui2017Correspondence} estimated a lower limit of 10~K for $T_{\rm d}$. Since, in this case, we are not aiming for a spatially resolved model, we leave these parameters unspecified to obtain average values for $T_{\rm d}$ and $N_{{\rm H}_2}$ treating the values given above as boundaries for the parameters in the fitting.

\subsubsection{Free-free emission (Thermal bremsstrahlung)}

Free-free emission is modeled following \cite{Stanislavsk2023}, namely:

\begin{equation}
I^{\rm ff}_{\nu} = 2.8 \times 10^{-16}~T_e~\bigg(\frac{\nu}{\mathrm{Hz}}\bigg)^2(1-e^{-\tau}) ~ 
\end{equation}
with 
\begin{equation}
    \tau = 3.014 \times 10^4 Z~\bigg(\frac{\nu}{\mathrm{MHz}}\bigg)^{-2}\bigg(\frac{\rm EM}{\mathrm{cm^{-6}~pc}}\bigg)~\bigg(\frac{T_e}{\mathrm{K}}\bigg)^{-1.5} g_{\rm ff}
\end{equation}
and 
\begin{equation}
g_{\rm ff} = \ln\bigg[\frac{49.55}{Z}   \bigg(\frac{\nu}{\mathrm{MHz}}\bigg)^{-1}\bigg] + 1.5 \ln\bigg(\frac{T_e}{\mathrm{K}}\bigg) .
\end{equation}
For simplicity, in this work we assume $Z=1$. The other parameters involved in the calculation are the electron temperature, $T_e$ and the emission measure, EM, treated as free parameters in the fitting.
Finally, the total thermal emission, $\nu F_\nu$, namely the energy flux, can be obtained by summing these two contributions to the specific brightness. We obtain $\nu F_{\nu}=\Omega_\gamma\nu~(I_{\nu}^{\rm d}+f_{\rm fill} I_{\nu}^{\rm ff}$), where $\Omega_\gamma$ is the solid angle corresponding to the gamma-ray extension and $f_{\rm fill}$ is the surface filling factor of the free-free emission within the circular area of radius $\theta_{\gamma}$, assuming that the free-free emission may come from a smaller projected region.

\subsection{Non-thermal emission}

Starting from the distribution of particles described in Sect.~\ref{subsec:primaries}, non-thermal emission was modeled using the \texttt{naima} software package, an open source code that incorporates all relevant cross-sections for radiative processes. We refer to \cite{Zabalza2015} for a complete description of the code. Below we briefly describe the radiative mechanisms and the parameter selection for each case.

\subsubsection{Pion-decay emission}

Protons and other nuclei exceeding an energy threshold of $\sim 300$~MeV generate pions -- unstable mesons that decay rapidly producing gamma rays and neutrinos -- when they interact with the ambient medium. These neutral products therefore carry information about the spectral distribution of the incident nuclei. The emissivity via this channel, once the particles spectrum is fixed, depends only on the target average gas density, $\langle n_{\rm H}\rangle$, being: 
\begin{equation}
F^{PP}_{\gamma} \propto \frac{\langle n_{\rm H}\rangle}{d^2} \ \int \frac{d \sigma_{pp\rightarrow \gamma}}{dE_{\gamma}}(E_p,E_\gamma) N_p(E_p) \,dE_p,
\end{equation}
where $d$ is the distance to RCW~38, recently confirmed to be $\sim 1600$~pc \citep{Zucker2019} and $d\sigma_{pp\rightarrow \gamma}/dE_{\gamma}$ is the cross-section of the process, for which we use the parametrization derived by \cite{Kafexhiu2014}. Assuming that the emitting region extends on the line of sight by $2R_\gamma$, an estimate of the  density is: $\langle n_{\rm H} \rangle = 2N_{{\rm H}_2}/(2R_{\gamma})$, where $N_{{\rm H}_2}$ is the column density derived from the fit of the dust thermal emission (see Sect.~\ref{sec:thermal}). 

\subsubsection{Bremsstrahlung}
\label{subsec:brems}

Non-thermal bremsstrahlung originates from the interaction of relativistic electrons with the ambient gas. The \texttt{naima} software package uses the cross section given by \cite{Baring1999}. The emissivity through this channel, once the electron population, and hence $K_{\rm ep}$, is fixed, depends solely on the ambient average density $\langle n_{\rm H}\rangle$,  which also determines the emissivity of the pion-decay channel. For what concerns $K_{\rm ep}$, no real constraints can be assumed {\it a priori}, as the ratio of accelerated electrons over protons depends on the microphysics of the shock, which is unknown for SCs. We simply assume in the fit that this ratio is smaller than 1, recalling that observations and simulations for the case of SNR shocks indicate values of $K_{\rm ep}$ ranging from 10$^{-4}$ to 10$^{-2}$ .

\subsubsection{Inverse Compton}\label{subsec:ic}

Inverse Compton scattering of relativistic electrons off background photon fields is a common mechanism for the production of gamma-ray emission. In this work, we accounted for three main photon fields: ({\it i}\/) the CMB; ({\it ii}\/) an infrared photon field, described by the grey-body that fits {\it Planck} and IRAS data ($T_{\rm d}= 40$~K and $U_{\rm rad}=10 $~eV~cm$^{-3}$); and ({\it iii}\/) the UV photon field of the massive stars. The latter was derived from the bolometric luminosity of the 29 OB star candidates identified by \cite{Wolk2006}, with $U_{{\rm rad,ts}}=\sum_i L_{{\rm bol},i} /(4 \pi c R^{2}_{\rm ts})\sim 290$~eV~cm$^{-3}$, assuming the same peak luminosity of 30,000~K for all stars. As it can be seen in Fig.~\ref{fig:eloss}, the dominant IC scattering in the energy range of our interest is that caused by IR photons. However, IC is subdominant in such a dust rich environment, since the medium density is  high enough to make pion and bremsstrahlung radiation mechanisms much more efficient than IC, at least in the MeV-GeV range. For this reason, including the IC contribution does not affect the fit and is added afterwards. 

\subsubsection{Synchrotron}
\label{subsec:sybc}
We model synchrotron emission resulting from the interaction of accelerated electrons with the ambient magnetic field. Synchrotron emission is expected to be significant from the radio to the X-ray band, although the highest-energy end of this component is determined by the magnetic field and the maximum energy of electrons. As the latter is determined by losses at the termination shock  in our model, the peak of synchrotron emission depends only on the magnetic field, $B$, in the emitting region, which in principle may be different from the magnetic field in the acceleration region. We do not know precisely where the emission region is located, either close to the TS or far away from it. A possibility is that electrons radiate more efficiently at the boundary between the hot wind bubble and the dense gas: here the magnetic field can be amplified due to MHD instabilities up to values even larger than $B_{\rm ts}$. Such a scenario is also supported by Zeeman line emission in a molecular cloud at the edge of the H\textsc{ii} region, where \cite{Bourke2001NewClouds} measured a magnetic field of $38~\mu$G.   
To be conservative, in the fit we assume a lower boundary for the magnetic field equal to $10~\mu$G. An upper limit could instead be derived by considering that the higher the magnetic field, the smaller the emission region. Therefore, a limit on the magnetic field can be derived by requiring that the travel time due to advection -- the dominant transport process for GeV electrons, responsible for the radio-synchtrotron emission -- is equal to the synchrotron loss time. We compute the advection time considering that particles must have traveled a distance corresponding to $R_{\gamma}$, and obtain $B\sim 180~\mu$G. Therefore, we round the upper limit for the magnetic field to 200~$\mu$G. 

Finally, we account for absorption of synchrotron radiation at radio frequencies due to free electrons in the plasma. The effect is more significant where free-free emission occurs, hence it depends on $f_{\rm fill}$. To account for this effect we consider the synchrotron brightness $I_{\rm syn} = F_{\rm syn}/\Omega_{s}$, defined as the flux per unit solid angle $\Omega_s$, and assume that a fraction $f_{\rm fill}$ of the area is affected by absorption while the rest is unabsorbed. We then obtain
\begin{equation}
\begin{split}
F_{\rm syn} = I_{\rm syn} e^{-\tau} f_{\rm fill}~\Omega_s +I_{\rm syn} (1-f_{\rm fill})~\Omega_s =&\\ 
= F^0_{\rm syn}[f_{\rm fill}(e^{-\tau}-1 )+1]&, 
\end{split}
\end{equation}
where $\tau$ is the opacity, and $F^0_{\rm syn}$ is the unabsorbed synchrotron emission that we derive by using  \texttt{naima}. 

\begin{table}[]
    \centering
    \begin{tabular}{l|ccl}
 \textbf{Parameter [Units]} & \multicolumn{2}{c}{\textbf{Value}}  &  \textbf{Reference} \\
& & & \\
\hline 
         $T_{\rm age}$ [Myr] &  \multicolumn{2}{c}{0.5} &  \cite{Wolk2006}\\
         $V_{\rm w}$ [km~s$^{-1}$] & \multicolumn{2}{c}{2600 }  & \cite{Pandey2024}\\
 $L_{\rm w}$ [erg s$^{-1}$]& \multicolumn{2}{c}{8$\times 10^{36}$} & \cite{Pandey2024}\\
 $d$ [pc]& \multicolumn{2}{c}{1600}  &\cite{Zucker2019} \\
 $\alpha_p$& \multicolumn{2}{c}{4.6} &\cite{Peron2024ThePopulation}\\
 $E_{p,{\rm max}}$ [TeV] & \multicolumn{2}{c}{100} & -- \\
 $E_{e,{\rm max}}$ [TeV]& \multicolumn{2}{c}{10} &  -- \\
 \hline 
  & \textbf{Min.} & \textbf{Max.} & \\
    $T_e$  [K]   & 5000  &12000 & \cite{Quireza2006}\\
     $\log$(EM/[cm$^{-6}$pc])&  4 & 8 & \cite{Khan2024}\\
      $f_{\rm fill}$   & 0.01& 1 & --\\
       $T_{d}$ [K]& 10  & 180  & \cite{Fukui2017Correspondence}\\
       $\log(N_{{\rm H}_2}$/[cm$^{-2}$])  & 20 & 22.5 & \cite{Wolk2002Discovery38} \\
       $\beta$ &  1& 2& --\\
       $ \log{K_{\rm ep}}$ & $-5$ & 0 & --\\
       $\log(B/[\mu$G])  & 1 & 2.3 & \cite{Bourke2001NewClouds}\\
       $\log{(N_{0,p}/ \mathrm{[eV^{-1}])}}$ & 35 & 40 & --  \\
\hline
         \end{tabular}

\vspace{0.2cm}
    \caption{Summary of the parameters used in the models and their priors. In the upper part of the table, we report parameters that are fixed in the fitting procedure; in the lower part of the table we list parameters that are left free to vary in the fit, with their corresponding prior boundaries.   }
    \label{tab:params}
\end{table}

\section{Results and discussion} 
\label{sec:result}

\begin{figure*}
    \centering
    \includegraphics[width=0.9\linewidth]{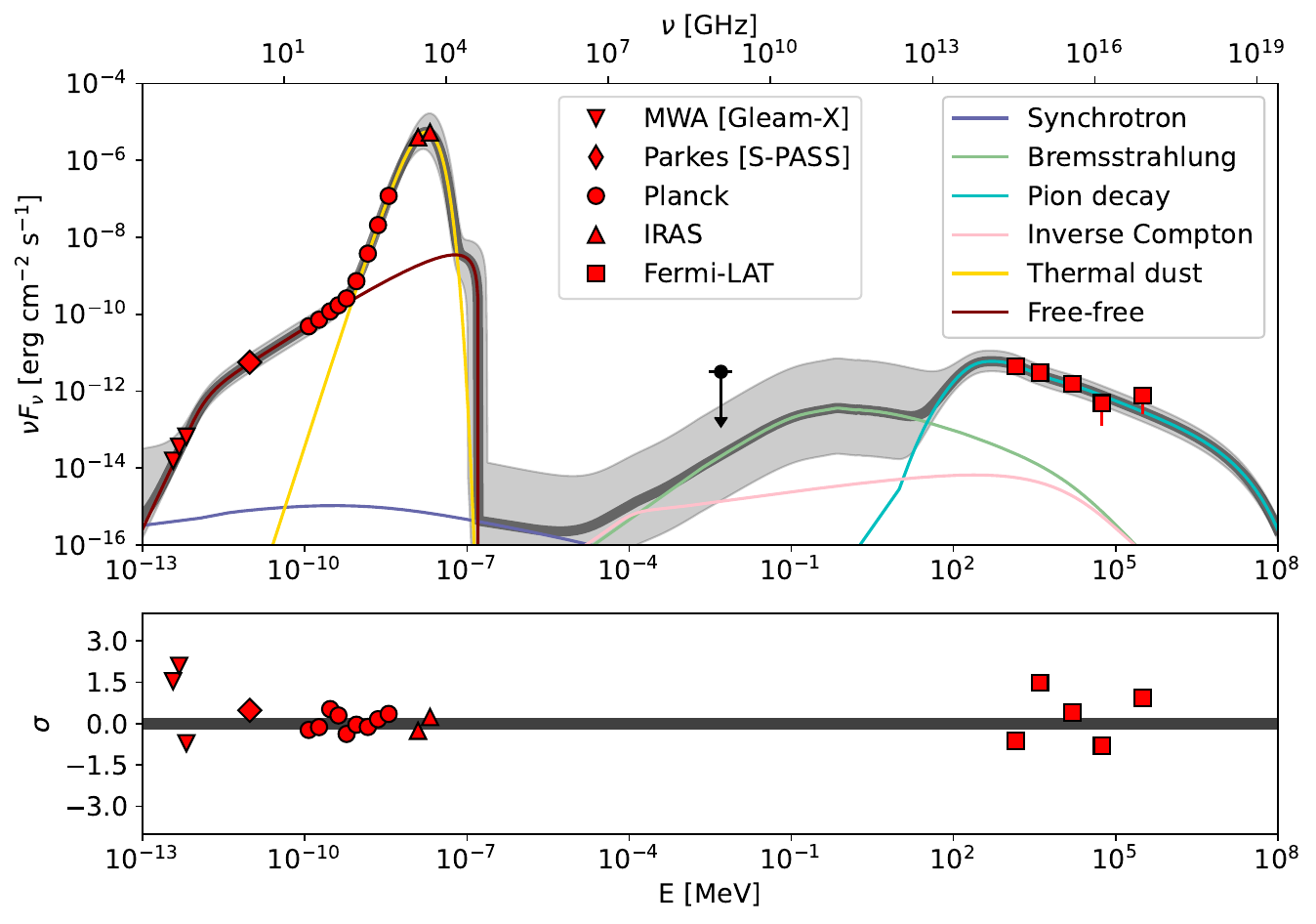}
    \caption{{In the top panel: the best-fit model for the broadband SED of RCW~38. Red points refer to the observations of the different instruments listed in the central inset, while the black upper limit indicates the level of the emission measured by Chandra, as explained in the text. The grey curve represent the sum of thermal and non-thermal components, listed in the figure legend, the grey area indicates the confidence levels corresponding to the 16th and 84th percentiles. In the lower panel: the residuals of the fit computed as explained in the text.}}
    \label{fig:fitted_sed}
\end{figure*}

\begin{figure*}
    \centering
    \includegraphics[width=1\linewidth]{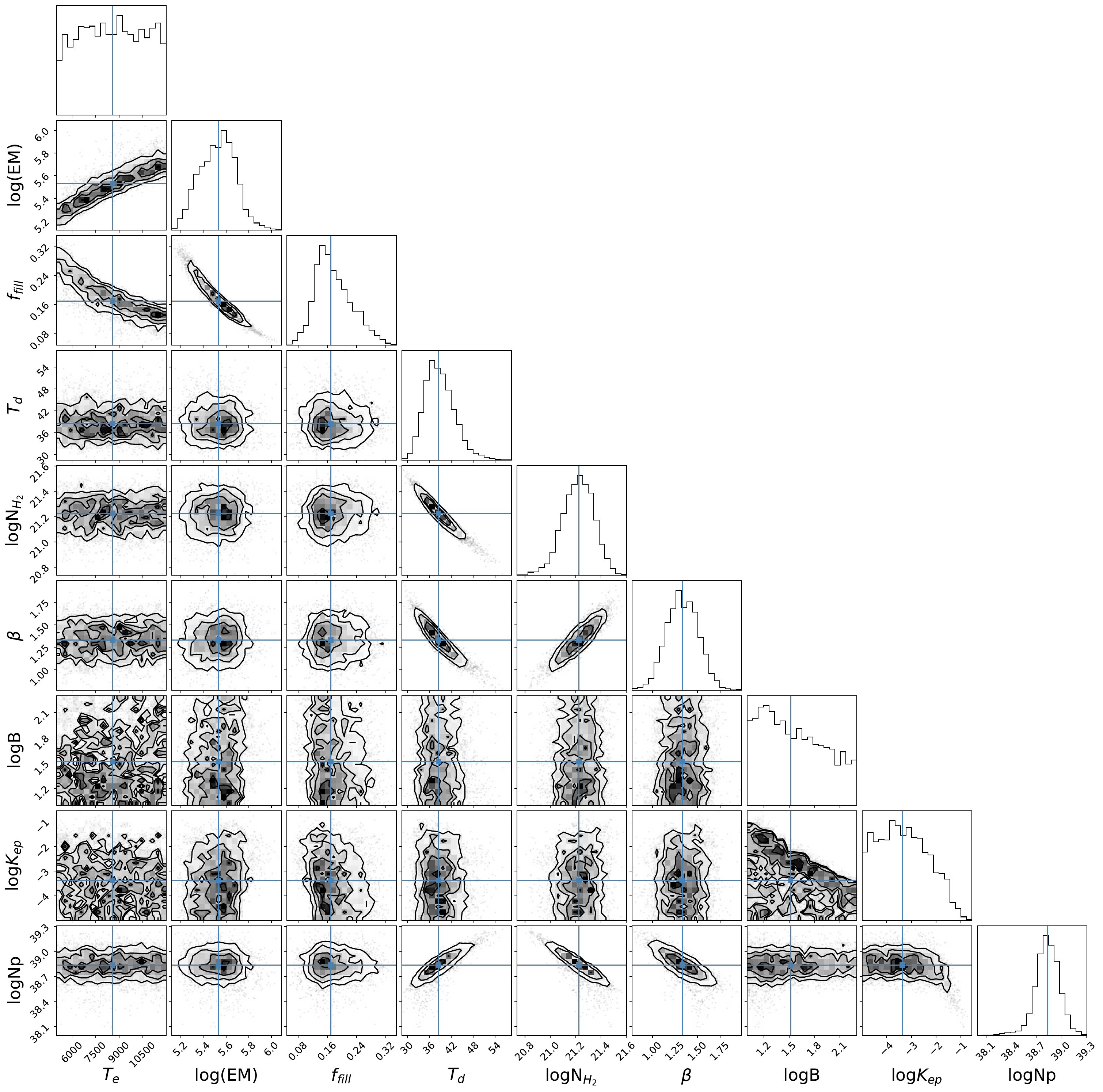}
    \caption{Corner plots for the fitted parameters in the model. }
    \label{fig:corner}
\end{figure*}

The final model, constructed by summing all the emission components described in Sect.~\ref{sec:model}, is fitted to the data, resulting in the spectral model shown in Fig.~\ref{fig:fitted_sed}, while the best-fit parameters with their uncertainties are listed in Table~\ref{tab:result_pars}. In the same figure, we also report the residuals, evaluated as $\sigma= ({F_{\rm obs}-F_{\rm mod}})/{\sigma_{\rm obs}}$, where $F_{\rm obs}$ and $F_{\rm mod}$ are the measured flux and the best-fit flux, while $\sigma_{\rm obs}$ is the error on each measured spectral point. The correlation between the different parameters is evaluated by the corner plots shown in Fig. \ref{fig:corner}. The corner plots provide quantitative indications on the quality of our fit, which, despite the large number of free-parameters, converges and constrains the parameters in a narrow interval of values. The good quality of the fit is testified 
by the small value of the reduced chi-square,  $\chi_\xi^2 = (N_{\rm obs}-\xi)^{-1}\sum^{N_{\rm obs}}_i ({F_{{\rm obs},i}-F_{{\rm mod},i}})^{2}/ \sigma^2_i $, where $N_{\rm obs}$ is the number of observed data points (20) and $\xi$ is the number of parameters (9). The value of $\chi^2_{\xi}$ is estimated to be 1.16.

\begin{figure}
    \centering
    \includegraphics[width=0.9\linewidth]{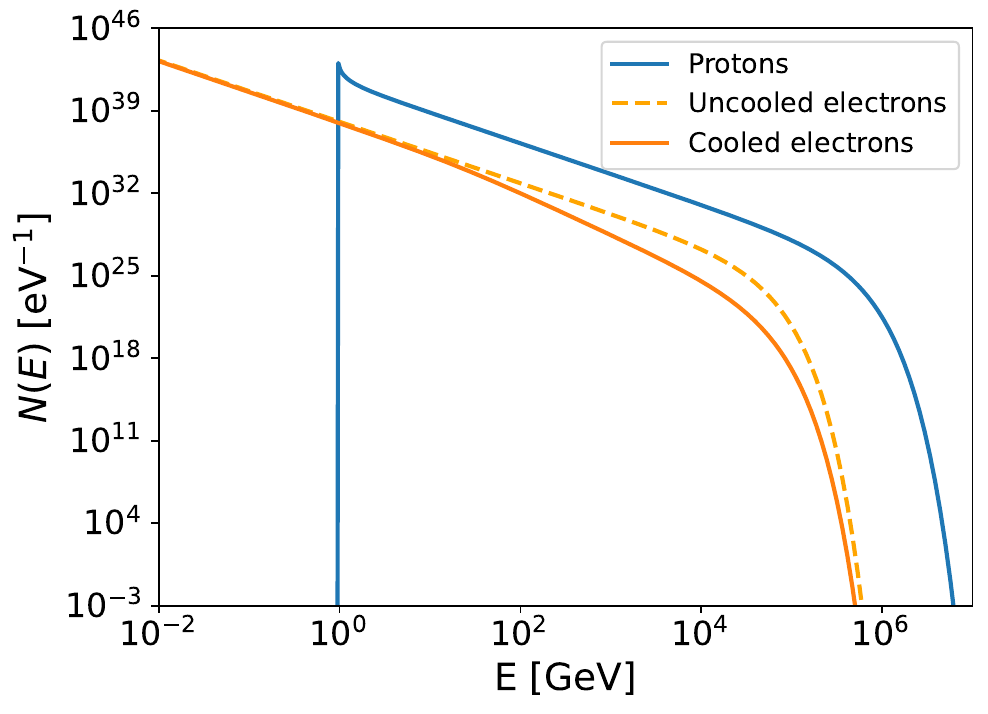}
    \caption{Electrons (orange) and protons (blue) spectral energy distribution (as a function of total energy of the particles) calculated including the terms of losses with the best-fit parameters listed in Table \ref{tab:result_pars}. For reference, we show also the un-cooled population of electrons, when the steepening due to losses is neglected. }
    \label{fig:pandelectrons}
\end{figure}

\begin{table}[]
    \centering
    \begin{tabular}{l|cc}
\textbf{Parameter [Units]}& \textbf{Best-fit value} & \textbf{Error} \\
 \hline 
    $T_e$ [$10^3$ K]    &  8.3&-2.3, +2.4\\
 $\log$(EM/[cm$^{-6}$pc])&  5.52&-0.18, +0.15\\
      $f_{\rm fill}$   & 0.17& -0.04, +0.07\\
       $T_{d}$ [K] &  38.7& -3.4, +4.1\\
       $\log(N_{{\rm H}_2}/[\rm cm^{-2}]$)  & 21.22  &-0.12, +0.11\\
       $\beta$ &  1.33& -0.16, +0.16\\
       $\log(B/[\mu$G])& 1.5& -0.4, +0.5\\
       $\log K_{\rm ep}$& -3.36  &-1.04, +1.2\\
       $\log{(N_p/ \mathrm{[eV^{-1}])}}$ & 38.84 & -0.14, +0.12\\
\hline
    \end{tabular}
    \vspace{0.2cm}
    \caption{Best-fit parameters and uncertainties corresponding to the 16th and 84th percentiles of the posterior distributions.}
    \label{tab:result_pars}
\end{table}
{
Synchrotron emission turns out to be sub-dominant in the radio regime compared to the free-free emission component, which well reproduces the set of data collected by MWA, Parkes and {\it Planck}. Moreover, the synchrotron-emission estimates should also be considered as upper limits, particularly at the lowest frequencies where additional suppression mechanisms such as the Razin–Tsytovich effect (RTE) may operate \citep[e.g.,][]{Ginzburg1965, Bracco2025}. Using our estimates of the electron density ($n_e$$\sim$$\sqrt{{\rm EM}/L} \approx 230$~cm$^{-3}$) and magnetic field ($B\sim30\, \mu$G), the corresponding lower cutoff frequency \citep[$20\, n_e/B$~MHz,][]{FeeneyJohansson2019} for the RTE is expected to be around~150~MHz. However, the best-fit magnetic field is consistent with the available Zeeman measurement \citep{Bourke2001NewClouds}, suggesting that suppression due to the RTE, if present, is likely negligible in the observed frequency range.

Modeling the radio data in terms of thermal emission imposes tight constraints on the accelerated electron population: although the electron fraction and the magnetic field are tightly correlated, we estimate that $K_{\rm ep}$ is of the order of $10^{-3}(B/30~\mu{\rm G})^{-1}$. Even in the minimum possible magnetic field, namely the interstellar one, of the order of $3~\mu \mathrm{G}$}, $K_{\rm ep}$ would be of the order of $10^{-2}$, implying that the bulk broadband emission that we measure cannot originate from non-thermal electrons. We instead support a hadronic origin of the emission detected by {\it Fermi}-LAT and a thermal origin of the X-ray emission detected by Chandra and Susaku. We further demonstrate in Appendix \ref{sec: syncX} that a non-thermal origin of the X-ray emission requires extreme assumptions, and therefore we disfavor this possibility.The resulting populations of electrons and protons that we derive from our fit are shown in Fig. \ref{fig:pandelectrons}.  

From the latter, we can derive the amount of energy retained by accelerated protons, which produce the detected gamma radiation. The best-fit value for the normalization at pivot energy of protons is $N_0 \approx 10^{39}~\mathrm{eV^{-1}}$, corresponding to a total energy in terms of protons of: 
 \begin{equation}\label{eq:power}
     W_p = \int^{\infty}_{1~\mathrm{GeV}} E \, N_p(E) \, dE \sim 8 \times 10^{47} \, \mathrm{erg}
\end{equation}
This can be compared with the luminosity  provided by the stellar winds, $L_{\rm w}$, to estimate the efficiency of particle acceleration, $\eta_{\rm CR}$. One important factor in this calculation is to compute the timescale at which the emission occurs; as the derived spectrum of particles is steep, protons of GeV energy dominate the energetic budget. As discussed in section \ref{subsec:primaries}, we consider  advection to be the dominant mechanism of particle transport and we consider as reference value $T_{adv}\sim 0.5~T_{age}$. The latter is comparable with the p-p interaction timescale: 

\begin{equation}
    T_{pp} = 1.6 \times 10^{13} \bigg( \frac{n}{100~ \mathrm{cm}^{-3}}\bigg)^{-1} \approx 0.5~\mathrm{Myr}
\end{equation}
These two timescales determine the production of gamma-rays, and therefore should be weighted in the calculation of the acceleration power. Normally, one should consider the bubble size and the propagation through it \citep[e.g.][]{Peron2025}, because only at the edge of the bubble the target density is large enough to produce detectable emission. In the specific case of RCW~38, however, the measured gas density is almost constant throughout the region \citep{Fukui2016}, allowing us to avoid to solve the transport equation through the bubble, as emission occurs in the immediate vicinity of the TS. Therefore we can easily estimate:

\begin{equation}
\eta_{\rm CR} = \frac{W_p}{L_{\rm w} \min [{T_{\rm adv},T_{\rm pp}}]} =  \frac{W_p}{L_{\rm w}  {T_{\rm adv}} }\approx 0.012.
\end{equation}

A much smaller value was derived through the rough estimate by \cite{Peron2024ThePopulation}, simply because a much larger value of $L_{\rm w}\sim10^{38}~\mathrm{erg~s^{-1}}$ was assumed.   
Other authors have instead obtained for RCW~38 a much larger value for the acceleration efficiency \citep{Pandey2024}, because they assumed a much shorter confinement time, dominated by diffusion with a Galactic diffusion coefficient. 
We assumed throughout the text that advection dominates up to energies $\sim$ 1 TeV, but we stress that if diffusion time were shorter than advection,the efficiency would increase accordingly. We can hence look at our value as a conservative lower limit to the acceleration efficiency of the cluster embedded in RCW~38. However, estimates of the diffusion coefficient with the magnetic field derived by our best fit (see Table \ref{tab:result_pars}) suggest  that in order to obtain a faster diffusion compared to advection at 10~GeV, we need for the Kraichnan case a coherence length larger than 0.1~pc, while for the Kolmogorov case a coherence length larger than 0.005~pc is sufficient. These quantities cannot be estimated a priori, however, the complexity of the environment with the presence of several small-size clumps, suggests that such a small coherence length may be possible. Furthermore, a too large value of $\eta_{CR}$ may produce an overproduction of the $^{22}$Ne isotope; as argued in \cite{Peron2024ThePopulation}, with average values of wind luminosities for OB stars in the Galaxy, the value of $\eta_{CR}$ should not exceed $\approx 10$\%. 
The derived lower limit for the efficiency, together with the confirmed hadronic nature of the emission, derived by the estimate of the value of $K_{\rm ep}$ that would be needed otherwise, supports the idea that winds of massive stars are able to contribute a small fraction of cosmic rays at Earth. 
\begin{figure}
    \centering
    \includegraphics[width=0.9\linewidth]{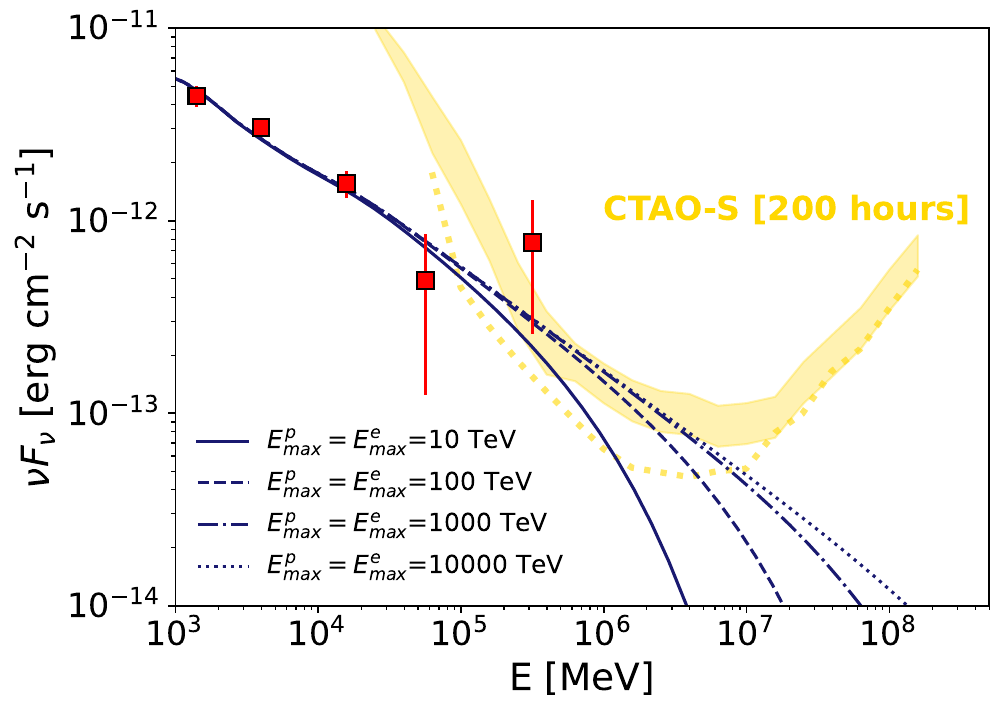}
    \caption{CTAO sensitivity of the Southern array compared with the SED of RCW~38 assuming different maximum energies of primary particles. The sensitivity is calculated for an exposure of 200 hours on-axis, using the prescription of \cite{Celli2024DetectionLHAASO},  namely requiring a minimum 10 events per bin with a significance of 5$\sigma$ over the background. The different golden curves represent the sensitivity for a point like source (dotted line) and for an extended source with radius ranging from 0.1$^{\circ}$ to 0.2$^{\circ}$ (area). } 
    \label{fig:CTAO}
\end{figure}

A question that remains unanswered is the maximum energy that particles can achieve in this early evolutionary phase. This critically depends on the properties of particle transport in the system. A relevant new development in this context concerns the effects of dust grains, largely abundant in the region of RCW~38, in triggering the growth of non-resonant streaming instability \citep{Gabici_dust}. This instability is known to lead to effective magnetic field amplification that can help reach very high energies \citep{bell04}.

As discussed throughout this work, the available data do not allow us to probe proton energies beyond a few TeV. This will be possible only with the next generation of ground-based observatories. The upcoming Cherenkov Telescope Array Observatory (CTAO), and in particular its Southern site (since RCW~38 is located in the Southern Sky) will have the necessary performance to evaluate the maximum energy of this system. We show this in Fig. \ref{fig:CTAO}, where we compare the flux of the region -- tuned on low-energy observations and computed under different assumptions on the maximum particle energy -- to the sensitivity of the CTAO southern site. The latter is shown for different assumptions for the size of the source \citep{Celli2024DetectionLHAASO}, and is derived by imposing a minimum detection threshold of 10 signal events with at least 5$\sigma$ significance in each energy bin. We find that, despite the severe worsening of the sensitivity of CTAO when observing extended regions, the observatory, provided a sufficiently long exposure, will be able to discriminate among the different proposed scenarios and hence to assess the maximum proton energy reached by this system. 

\section{Conclusions and outlook}
\label{sec:conclusion}

In summary, we studied the broadband emission -- from radio to gamma rays --  originating from the H\textsc{ii} region RCW~38, which hosts a very young massive star cluster, and drew conclusions on the nature of the emission in different energy bands. We confirmed the thermal origin of the radio and X-ray emission, and the hadronic nature of gamma-ray non thermal radiation. Based on this new, more extended modeling, we refined the estimate of the acceleration efficiency, finding that at least $\sim 1\%$ of the wind power is used to energize CR protons {responsible for the gamma-ray emission}. Finally, we provided the first model-independent estimate of the electron/proton ratio accelerated in young stellar systems, namely $K_{\rm ep}\simeq 10^{-3} (B/30{\rm \mu G})^{-1}$, finding that it is consistent with the value inferred for young SNRs. Interestingly, it has been noted for SNRs that smaller values of $K_{\rm ep}$ are recorded for younger systems, when the shock velocity is higher, while higher values are inferred for middle aged SNRs \cite[see discussion in section~6.1 in][]{Morlino-Celli:2021}.The velocity of the winds in RCW~38 is in the same range of velocity of young SNRs (of the order of 3000 km s$^{-1}$). If the electron injection efficiency in the DSA mechanism is determined by the velocity of the shock \cite[see e.g.][]{Ghavamian13, Raymond23}, this might be the reason why the obtained value of $K_{\rm ep}$ is similar for the two cases. This finding supports the idea that the acceleration physics is similar in the two classes of sources and, more generally, for shocks of comparable speed.

Our results are apparently at odds with the conclusions recently derived on the Cygnus~OB2 stellar association and on the Westerlund~1 SC. Recent analyses by \cite{HaererWd1,HaererCygnus} attribute the bulk of the emission to leptonic processes, mainly due to lack of a sufficient amount of gas to interpret the emission as hadronic. This interpretation, however, implies a much larger $K_{\rm ep}$.
On the other hand, it should be noted that the conditions of particle acceleration in 
environments where the most massive stars have already exploded as SNe may be different from the conditions found in very young embedded clusters, like RCW~38, where SNe should not have occurred yet. Although this requires further investigation, if confirmed, the scenario would point to  different regimes of acceleration in terms of hadrons vs. leptons for SCs in different evolutionary stages.

The desirable advent of a sensitive instrument in the 1--100 MeV regime, like newASTROGAM or AMEGO-X, could further help 
to discriminate between hadronic and leptonic emission and constrain the $K_{\rm ep}$ ratio \citep[as discussed, e.g., in][]{Peron2026} 
in a number of Galactic CR accelerators.
Finally, in this work the maximum acceleration energy remains unconstrained due to the lack of data beyond a few hundreds of GeV. In this regard, we discussed the role of next-generation ground based instruments, like ASTRI-MiniArray and CTAO, which will be essential to test the high-energy limit achievable in these stellar systems and ultimately clarify their role as possible hadronic PeVatrons. 
 
\begin{acknowledgements}

We acknowledge discussions with Enrico Peretti, Niccol\`o Bucciantini, and Rino Bandiera. We thank Chenoa D. Tremblay and John M. Dickey for sharing first generation MWA data at 114 MHz of RCW 38 that were subsequently updated with the GLEAM-X survey. This work has made use of S-band Polarisation All Sky Survey (S-PASS) data. Some of the results in this paper have been derived using the {\tt healpy} and {\tt HEALPix} packages. The work of GP and AB is supported by the INAF Astrophysical Fellowship (IAF) initiative that we both acknowledge, while the work of GM is partially supported by the INAF Theory Grant 2024 {\it Star Clusters As Cosmic Ray Factories II}.

\end{acknowledgements}

\bibliographystyle{aa} 
\bibliography{references,references-2} 

\begin{appendix}

\section{Comparison of different {\it Fermi}-LAT analyses}
\label{sec: FermiCFR}
Throughout this paper we refer to the results of the analysis of RCW~38 conducted by \cite{Peron2024ThePopulation} (hereafter GP24),  however, later analyses of the same region were performed by \cite{Pandey2024} (PP24) 
 and by \cite{Ge2024RCW38} (TTG24). Here we briefly summarize the consistency of these independent analyses.
\begin{table}[!h]
    \centering
    \begin{tabular}{lccc}\toprule
         &   $\mathbf{\theta} $& $\mathbf{\Gamma}$&  \textbf{Energy range}\\
 & [$^\circ$]& &[GeV]\\\midrule
        GP24& 0.21$\pm$0.04
        
     & 2.56 $\pm$0.05 &1--1000\\
 PP24& 0.24 $\pm$ 0.04&  2.34 $\pm$ 0.04& 0.1--500 \\
 TTG24& 0.23 $\pm$ 0.02 &  2.44 $\pm$ 0.03&0.1--500\\ \bottomrule\end{tabular}
    \caption{Comparison of parameters used in different analysis to describe the region of W44}
    \label{tab:cfr_analysis}
\end{table}

\begin{figure}[!h]
    \centering
    \includegraphics[width=1\linewidth]{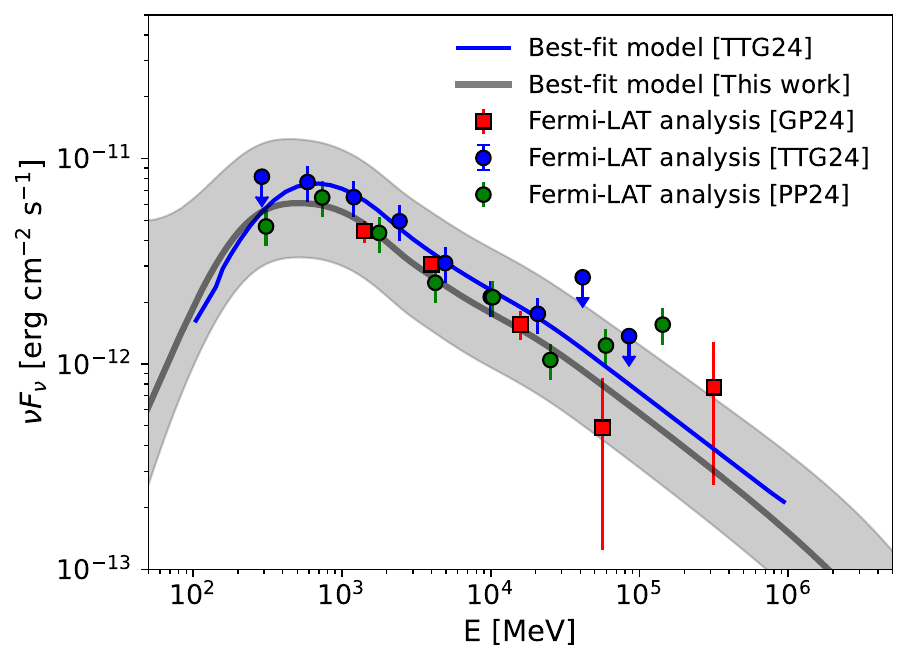}
    \caption{Comparison of the SEDs derived in different studies, as described in the text. The blue curve is the best fit model derived by \cite{Ge2024RCW38}, while the gray curve and the gray shaded area represent the best-fit model and its confidence levels as derived in this work. }
    \label{fig:cfr_SEDs}
\end{figure}

As one can see from Table \ref{tab:cfr_analysis}, small differences are reported in the spectral index, probably due to the different sampled energy range: both PP24 and TTG24 consider photons in the sub GeV range, where the emission flattens due to the characteristic pion-bump feature. Their modeling as a simple power-law result in a harder spectral index. As one can see in Fig. \ref{fig:cfr_SEDs}, the SEDs are in general compatible and all data-sets agree with the model we derived in this work. The two best-fit models differ by about 20\%. We therefore conclude that the small systematic differences that are found among different studies do not affect the conclusions of this work. 
 
\section{Non-thermal origin of X-ray diffuse emission}
\label{sec: syncX}
We investigate whether the diffuse emission observed by Chandra and Suzaku \citep{Fukushima2023}, that we assumed to be of thermal origin throughout the paper, could rather be interpreted as non thermal emission, namely synchrotron, bremsstrahlung, or inverse Compton emission of relativistic electrons accelerated inside RCW~38. We want to verify if such a picture, suggested by some authors in the past \citep[e.g.][proposed a synchrotron origin due to multi-TeV electrons]{Wolk2002Discovery38} can explain the entire set of broadband data. 
The photon index that is attributed to the non-thermal flux by these authors is $\Gamma = 1.7\pm 0.4$ for the inner ($<2'$) part and $1.8\pm 0.1$ for the outer ($2'-5'$) part. We recall that, for both synchrotron and IC emission, the spectral index of electrons, $\beta$, ($\beta\approx \alpha_e-2$, for $E_e\gg m_ec^2$) is linked to the photon index $\Gamma$ by $\beta=2\Gamma-1$.  Therefore, using the best fit values, one obtains a $\beta$ of 2.4 and 2.6 (corresponding to $\alpha_e=4.4$ and 4.6 in terms of momentum) for the inner and outer part respectively. These values are similar to what found for protons, and require that the electron spectrum is not appreciably affected by losses, if the same injection spectrum is assumed for both species. 
Losses are especially important in the energy range relevant for X-ray synchrotron (a few tens of TeV) as one can clearly see from Figure~\ref{fig:eloss}. A possible way out is to assume that the emission is fully leptonic, so that the spectral index of protons would be left unconstrained. In this case, if the X-ray spectrum is to be explained as synchrotron emission, the spectral injection slope should be $s=\alpha_e-1$ between 3.4 and 3.6, difficult to reconcile with DSA, which predicts injection spectra with index $s\gtrsim 4$, but in principle not forbidden. 

\begin{figure*}
    \centering
   \includegraphics[width=0.5\linewidth]{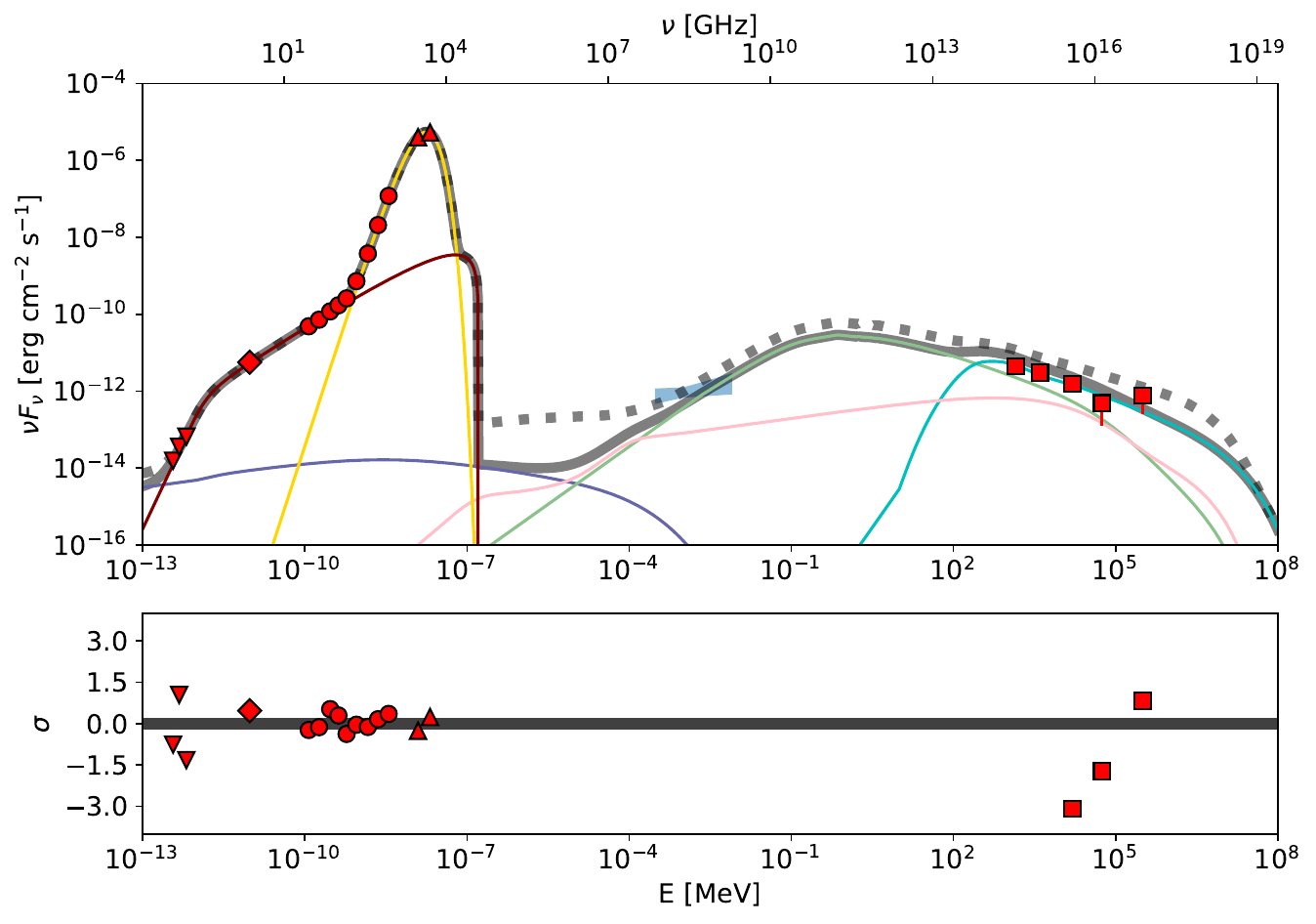}\includegraphics[width=0.5\linewidth]{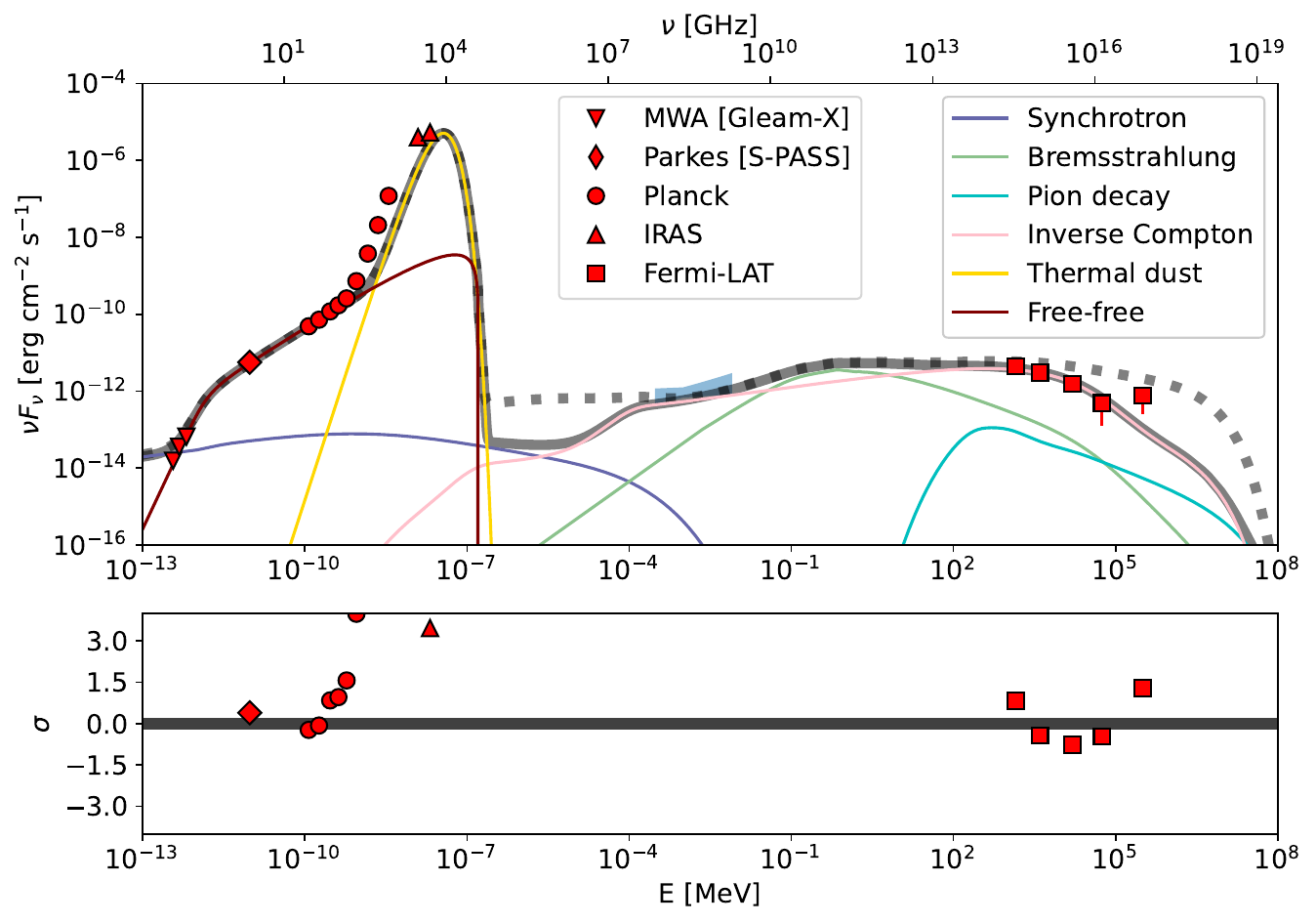}
    \caption{Broadband modeling of the emission under the assumption of non-thermal origin of the X-ray flux. The latter is represented by a blue butterfly, which accounts for the uncertainties in the flux normalization and slope as in \cite{Fukushima2023}. On the left panel, the bremsstrahlung component is increased, compared to the nominal model presented in the main text, by increasing $K_{ep}$ to 0.05. Alongside, the magnetic field is decreased to 10 $\mu$G to avoid that the synchrotron emission overshoots the data. In the right panel the IC component is increased by increasing  $K_{ep}$ to 0.2, the magnetic field is decreased to 10 $\mu$G, and the gas column density to $3\times10^{19}$cm$^{-2}$; to adjust the normalization of thermal dust, its temperature is increased to 80 K. In both plots the grey solid curve represents the total emission, while the dotted grey curve is the total emission, when neglecting the electron cooling.  }
    \label{fig:ic_prova}
\end{figure*}

The best-fit flux for synchrotron that we found in the main text (see Fig \ref{fig:fitted_sed}), however, is a few order of magnitudes below the measured flux in the X-ray band. Still this is the maximum level allowed by the radio measurements.
Increasing the synchrotron component either by increasing the electron number density or the magnetic field would unavoidably result in an overestimation of the radio data. In addition, increasing the magnetic field would make the energy losses more severe, decreasing the maximum energy below the threshold required for the electrons to emit X-ray photons. For this reasons, the synchrotron scenario is here excluded. 
Concerning the bremsstrahlung, we can reach a value compatible with the X-ray flux by increasing $K_{\rm ep}$ up to 0.05 (see left panel of Figure~\ref{fig:ic_prova}). However,at energies below the electron rest-mass, the bremsstrahlung spectrum has a typical fixed shape $\sim E^{-1}$, which is not compatible with the derived slopes in the X-ray band. The agreement with the gamma-ray data also becomes less evident, as it appears from the residuals. For this reason, we can safely exclude also a bremsstrahlung origin of the X-ray emission.
Finally, as far as the possibility of an Inverse Compton scattering origin is concerned, we can match the X-ray flux by increasing again $K_{\rm ep}$ but lowering at the same time the gas density to decrease the bremsstrahlung contribution.However, this comes with a price, as the gas density enters not only the Bremsstrahlung and the Pion decay emission, but also the emission flux of the thermal dust. Lowering the gas density lowers the thermal dust component, hence it is necessary to increase the dust temperature to reach the IR peak. In the right panel of Figure~\ref{fig:ic_prova} we have performed a fit assuming a dominant IC contribution: the curves have been obtained using $T_d$=80~K, $K_{\rm ep}=0.2$, $N_H =3\times 10^{19}$ cm$^{-2}$ and $B=10\mu$G. One can see that the residuals in the radio-microwave bands worsen significantly. A self-consistent model is hence not easy to achieve when accounting for all the observables. Furthermore, the obtained spectrum of IC is somewhat flatter than what observed in the GeV band, producing a worsening of the fit in this energy band, too. In addition, the required value for the column density is more characteristic of diffuse gas, rather than of dense region like the molecular cloud complex in which the cluster is embedded.

We therefore conclude that a non-thermal origin for the X-ray emission should be disfavored, and argue for a thermal origin of the X-ray emission.

\section{Radio data analysis complements }
\label{sec:appendix_radio}

In this Appendix, we summarize the main characteristics of the various datasets presented in Sect.~\ref{sec:obs} (see Table~\ref{tab:maps}), except for {\it Fermi}-LAT \citep[i.e.,][and references therein]{Peron2024ThePopulation}  and describe the methodology to compute and remove the background emission of RCW~38. This methodology was applied at all frequencies but to the gamma-ray data. 

\begin{table}[!h]
    \centering
    \footnotesize
    \begin{tabular}{c|c|c|c|c}
        {\bf Telescope} & {\bf Frequency}&  {\bf Bandwidth} & {\bf Beam} & {\bf Pixel}  \\
        & $\nu_0$ & $\Delta \nu/\nu_0$& $\theta_{\rm max}\times \theta_{\rm min}$ &  {\bf size}\\
        & [GHz] & [\%]  &[arcmin$^2$]& [arcmin]\\
    \hline 
    \hline 
    & & & \\
       MWA  &  0.088 & 35 & $2.4\times 1.9$ & 0.4\\
       MWA  &  0.118 & 26 & $1.8\times 1.4$ & 0.3\\
       MWA  &  0.155 & 19 & $1.4\times 1.1$ & 0.2\\
       & & & \\
    \hline
    & & & \\
       Parkes & 2.3 & 7& $8.9\times 8.9$ & 3.4 \\
       & & & \\
    \hline 
    & & & \\
       {\it Planck} & 30 & 20 &  $32.3\times 32.3$ & 3.4\\
       {\it Planck} & 44 & 20 & $27\times 27$& 3.4\\
       {\it Planck} & 70 & 20 & $13.3\times 13.3$ & 3.4\\
       {\it Planck} & 100  & 33& $9.7 \times 9.7$ & 1.7\\
       {\it Planck} & 143  & 33 & $7.3\times 7.3$& 1.7\\
       {\it Planck} & 217  & 33& $5\times 5$& 1.7\\
       {\it Planck} & 353  & 33& $4.9\times 4.9$& 1.7\\
       {\it Planck} & 545 &  33& $4.8\times 4.8$& 1.7\\
       {\it Planck} & 857 &  33& $4.6\times 4.6$& 1.7\\
       & & & & \\
    \hline 
    & & & \\
       {IRAS-4} & 2997 & 37& $4.3\times 4.3$ & 3.4\\
       {IRAS-3} & 4997 & 75& $4\times 4$ & 3.4\\
       & & & \\
       
    \hline    
    \end{tabular}
    \vspace{0.2cm}
    \caption{Observational properties of the multi-wavelength datasets used in the analysis.}
    \label{tab:maps}
\end{table}
To estimate the integrated flux of RCW~38 while accounting for local background emission, we adopted an aperture photometry approach based on a fixed source region and a surrounding annulus. For each map, we defined a circular aperture (hereafter, disk) of radius $\theta = \theta_{\gamma}$ centered on the source position. The signal within the disk was integrated after subtracting a locally estimated background level. The background was determined by computing the median of the pixel intensity distribution within an external annular region extending from $\theta$ to a variable outer radius $\theta_{\mathrm{out}}$. The use of the median provides a robust estimate of the background level, minimizing the impact of outliers and residual compact structures. By progressively increasing $\theta_{\mathrm{out}}$ between 0.3$^{\circ}$ and 3$^{\circ}$, we constructed a radial profile of the background-subtracted integrated flux and identified a plateau corresponding to a stable background estimate within 1\%. The final flux at a given frequency $\nu$ was obtained as
\begin{equation}
F_{\nu} = \sum_{i \in \mathrm{disk}} \left( I_{\nu,i}- I_{\nu,\mathrm{bg}} \right) \, \Omega^{\mathrm{pix}}_{\nu,i},
\end{equation}
where $I_{\nu,i}$ is the specific intensity in pixel $i$, $I_{\nu,\mathrm{bg}}$ is the median background level measured in the annulus, and $\Omega^{\mathrm{pix}}_{\nu,i}$ is the solid angle of the pixel, whose sizes are listed in Table~\ref{tab:maps}. The integrated flux density $F_\nu$ was then converted from Jy to erg\,s$^{-1}$\,cm$^{-2}$ by multiplying by the corresponding frequency, i.e., computing $\nu F_\nu$. This method is independent of the underlying pixelization scheme and was consistently applied to both {\tt HEALPix} and cartesian maps. 

\end{appendix}

\end{document}